\documentstyle[onecolumn]{mn}
\newif\ifAMStwofonts
\def\arcdeg{\hbox{$^\circ$}}
\def\arcmin{\ifmmode^\prime\;\else$^\prime\;$\fi}

\ifoldfss
  \ifCUPmtlplainloaded \else
    \NewTextAlphabet{textbfit} {cmbxti10} {}
    \NewTextAlphabet{textbfss} {cmssbx10} {}
    \NewMathAlphabet{mathbfit} {cmbxti10} {} % for math mode
    \NewMathAlphabet{mathbfss} {cmssbx10} {} %  "   "    "
  \fi
  \ifAMStwofonts
    \ifCUPmtlplainloaded \else
      \NewSymbolFont{upmath} {eurm10}
      \NewSymbolFont{AMSa} {msam10}
      \NewMathSymbol{\upi}     {0}{upmath}{19}
      \NewMathSymbol{\umu}     {0}{upmath}{16}
      \NewMathSymbol{\upartial}{0}{upmath}{40}
      \NewMathSymbol{\leqslant}{3}{AMSa}{36}
      \NewMathSymbol{\geqslant}{3}{AMSa}{3E}

      \let\leq=\leqslant 
       
    \fi
  \fi
\fi % End of OFSS

\ifnfssone
  \newmathalphabet{\mathit}
  \addtoversion{normal}{\mathit}{cmr}{m}{it}
  \addtoversion{bold}{\mathit}{cmr}{bx}{it}
  \newmathalphabet{\mathbfit} % math mode version of \textbfit{..}
  \addtoversion{normal}{\mathbfit}{cmr}{bx}{it}
  \addtoversion{bold}{\mathbfit}{cmr}{bx}{it}
  \newmathalphabet{\mathbfss} % math mode version of \textbfss{..}
  \addtoversion{normal}{\mathbfss}{cmss}{bx}{n}
  \addtoversion{bold}{\mathbfss}{cmss}{bx}{n}
  \ifAMStwofonts
    \ifCUPmtlplainloaded \else
      \UseAMStwoboldmath
      \makeatletter
      \new@mathgroup\upmath@group
      \define@mathgroup\mv@normal\upmath@group{eur}{m}{n}
      \define@mathgroup\mv@bold\upmath@group{eur}{b}{n}
      \edef\UPM{\hexnumber\upmath@group}
      \new@mathgroup\amsa@group
      \define@mathgroup\mv@normal\amsa@group{msa}{m}{n}
      \define@mathgroup\mv@bold\amsa@group{msa}{m}{n}
      \edef\AMSa{\hexnumber\amsa@group}
      \makeatother
      \mathchardef\upi="0\UPM19
      \mathchardef\umu="0\UPM16
      \mathchardef\upartial="0\UPM40
      \mathchardef\leqslant="3\AMSa36
      \mathchardef\geqslant="3\AMSa3E

      \let\leq=\leqslant 

    \fi
  \fi
\fi % End of NFSS release 1

\ifnfsstwo
  \DeclareMathAlphabet{\mathbfit}{OT1}{cmr}{bx}{it}
  \SetMathAlphabet\mathbfit{bold}{OT1}{cmr}{bx}{it}
  \DeclareMathAlphabet{\mathbfss}{OT1}{cmss}{bx}{n}
  \SetMathAlphabet\mathbfss{bold}{OT1}{cmss}{bx}{n}
  \ifAMStwofonts
    \ifCUPmtlplainloaded \else
      \DeclareSymbolFont{UPM}{U}{eur}{m}{n}
      \SetSymbolFont{UPM}{bold}{U}{eur}{b}{n}
      \DeclareSymbolFont{AMSa}{U}{msa}{m}{n}
      \DeclareMathSymbol{\upi}{0}{UPM}{"19}
      \DeclareMathSymbol{\umu}{0}{UPM}{"16}
      \DeclareMathSymbol{\upartial}{0}{UPM}{"40}
      \DeclareMathSymbol{\leqslant}{3}{AMSa}{"36}
      \DeclareMathSymbol{\geqslant}{3}{AMSa}{"3E}

      \let\leq=\leqslant 

    \fi
  \fi
\fi % End of NFSS release 2

\ifCUPmtlplainloaded \else
  \ifAMStwofonts \else % If no AMS fonts
    \def\upi{\pi}
    \def\umu{\mu}
    \def\upartial{\partial}
  \fi
\fi

\title{The Star Formation Histories of Two Northern LMC Fields}
\author[A. E. Dolphin]
       {A. E. Dolphin  \\
        National Optical Astronomy Observatories, P.O. Box 26732, Tucson, AZ 85726, USA}
\date{Accepted .
      Received ;
      in original form 1999 July 4}

\pagerange{\pageref{firstpage}--\pageref{lastpage}}
\pubyear{1999}

\begin{document}

\maketitle

\label{firstpage}

\begin{abstract}
Ground-based $UBV$ photometry of two fields in the northern disk of the LMC are presented.  A distance modulus of $(m-M)_0 = 18.41 \pm 0.04$ and an extinction of $A_V = 0.30 \pm 0.05$ has been calculated for these fields.  The measurable star formation history of the LMC began no more than 12 Gyr ago with a strong star forming episode with [Fe/H] = -1.63 $\pm$ 0.10 that accounted for approximately half (by mass) of the LMC's total star formation in the first 3 Gyr.  The data does not give accurate star formation rates during intermediate ages, but there appears to have been a recent increase in the star formation rate in these fields, beginning approximately 2.5 Gyr ago, with the current metallicity in the region being [Fe/H] = -0.38 $\pm$ 0.10.  The two fields have had very similar star formation rates until 200 Myr ago, at which point one shows a large increase.
\end{abstract}


\section{INTRODUCTION}
The LMC is the Milky Way's best-studied companion galaxy, a result of both its proximity ($\sim$50 kpc) and high galactic latitude.  Its stellar content is easily accessible from ground-based telescopes, allowing 1m-class telescopes to obtain effectively deeper data than the HST can in more distant Local Group members such as WLM and LGS 3.

The global properties of the LMC are well-studied, with a total mass of $3.5\times10^9 M_\odot$ and neutral gas mass of $5.2\times10^8 M_\odot$ (15\% of the total), including HeI, determined by Kim et al. (1998b).  Kim et al. (1998a) also overlaid an H$\alpha$ map over their HI map, showing the highest concentration of H$\alpha$ in the 30 Doradus region.

As a result of the LMC's proximity, much work has already been done on this subject, much of which is well summarized in a review by Olszewski, Suntzeff \& Mateo (1996).  The majority of star clusters seem to have mostly formed in the past 4 Gyr, with the remaining clusters formed over 12 Gyr ago.  The clusters also show a rise in metallicity over the galaxy's history, from -1.8 15 Gyr ago to -0.5 at present, both values with a spread of $\pm$0.5 dex.  Whether the formation times and metallicities of the clusters trace those of the global star formation, however, is uncertain.

The distribution of ages of the field stars is more difficult to measure.  An early attempt at studying the disk stars was made by Butcher (1977), who identified a break in the main-sequence LF as indicating that most of the stars have been formed in the past 3-5 Gyr.  A study of the star formation history of the LMC field stars using comparisons with synthetic CMDs was made by Bertelli et al. (1992), which confirmed the Butcher (1977) results.  It was determined that a burst of star formation activity began some 3-5 Gyr ago, with a star formation rate as high as ten times the rate at older ages.  These results confirm the expectations from the cluster studies.

Recently, however, HST results have brought this seemingly firm picture into question.  Holtzman et al. (1997) take HST data for the same field studied by Butcher (1977) and Bertelli et al. (1992), and obtain a different result.  They find that the star formation rate increase happened more recently (2-3 Gyr ago) and that, more importantly, the rate increase was only approximately a factor of 3.  This result was confirmed with the addition of ground-based data of the same field and surrounding regions by Stappers et al. (1997), and Geha et al. (1998) find that this star formation history is also correct for two other outer LMC fields.  It has also been hinted that the star formation rates may have begun to decrease during the past Gyr.

Olsen (1999) studied the HST field surrounding NGC 1754, which also lies in the northern part of the disk, as well as the fields surrounding four globular clusters in the bar.  Using a method much like that used in this work, he found that the Holtzman et al. (1997) results were qualitatively correct, but that the star formation enhancement during the past 3 Gyr has not been a uniform star formation rate.  Olsen (1999) also finds evidence of the decrease in star formation rate during the past 0.5 Gyr.  Similarly, Holtzman et al. (1999) find that the field star formation history differs from the observed star cluster formation history, and have results consistent with those of Olsen (1999).

For the upper main sequence, Holtzman et al. (1997) find a metallicity of between [Fe/H]=-0.7 and -0.4, while Geha et al. (1998) find the [Fe/H]=-0.4 isochrone to fit the best.  The abundances of the old stars are less constrained, with both groups finding that the red giant branch appears to have an old population with [Fe/H]=-1.7 but could have the upper RGB dominated by younger (less than 2 Gyr) stars with [Fe/H]=-1.3.

This study will present data for two fields near the Constellation III region, in the northern part of the LMC disk, as well as a star formation history analysis.

Goals for this study will be:
\begin{itemize}
\item Determining a photometric distance to the LMC, and comparing with other measurements.
\item Dating the oldest stars in the galaxy, and determining of star formation rates since that time.  Special importance will be attached to the 3-5 Gyr range, and the ratio of recent to intermediate age star formation rates, as all previous ground-based work has supported one interpretation, while all HST-based work has supported another.
\item Determining the chemical enrichment history, and whether the red giant branch contains a significant number of young stars.
\item Comparing the two fields for differences in recent star formation rates.
\end{itemize}

\section{DATA}

\subsection{Observations}

$UBV$ images of five fields inside Constellation III and two fields outside the region were obtained by Deidre Hunter at the Cerro Tololo Inter-American Observatory (CTIO) 26, 27, 30, and 31 January 1993 with a 2048$\times$2048 Tektronix CCD on the 0.9 m telescope.  Conditions were photometric (except as mentioned below), with seeing typically 1-2\arcsec.  The fields outside the Constellation will be the ones studied here, referred to as the SE and NW fields, with positions given in Table \ref{tab-4-radec}.  Additionally, Figure 1 of Dolphin \& Hunter (1997) shows rough sizes and locations of the seven fields.  Images in each filter were obtained with at least two integration times to give greater dynamic range, as well as for cosmic ray cleaning.  Standard stars were also observed, chosen from the lists of Graham (1982) and Landolt (1973; 1983).

\begin{table*}
\caption{Positions of the LMC Fields, epoch 2000}
\begin{tabular}{lrr}
Field & RA & Dec \\
\hline
NW & 5$^h$ 10$^m$ 02.1$^s$ & $-$66\arcdeg\ 36\arcmin\ 24\arcsec\ \\
SE & 5$^h$ 40$^m$ 12.7$^s$ & $-$67\arcdeg\ 11\arcmin\ 46\arcsec\ \\
\end{tabular}
\label{tab-4-radec}
\end{table*}

Image reduction was done normally, although there were some problems with the images.  The CCD was non-linear, with an error of 0.015 magnitudes per magnitude, which was corrected with a correction determined by Alistair Walker (private communication).  Some standard star images showed a step-function in the bias, for which only the part of the images with the normal bias was used so that no additional uncertainty was added to the photometry.  This bias problem was not observed in the data frames.
Additionally, all images with integration times of $\leq$5 seconds were corrected for the finite time required for the shutter to open and close.  Bright but unsaturated stars also had additional structures in their PSFs that could not be accounted for by PSF fitting, so those stars were not included in the photometry.

An iterative fit for the transformations to standard filters was made, with a zero point, airmass term, and colour term for each filter.  We required that the colour term and zero point were constant for the four nights, allowing only the extinction to vary.  In the data for the second night, there were two $U$ images and one $B$ image for which the standard star photometry was 0.15 magnitudes fainter than the rest.
Because the images were interleaved and because the offset was exactly the same in each case, the offset is unlikely to be due to clouds, although no other problems were obvious.  All other images agreed with the brighter level, so those three images were removed from the solution.  The rms's of the final solutions were 0.05 in $U$ and 0.02 for $B$ and $V$.

\subsection{Reduction}

The reduction procedure for this data is detailed in Dolphin \& Hunter (1997).  Only a brief outline is given here.

DAOPHOT (Stetson 1987) in IRAF was used to obtain photometry, following the procedure described by Massey \& Davis (1992).  Because of significant PSF variations in the chip, we were forced to use a quadratically varying PSF solution, which created minor differences from the standard DAOPHOT photometry recipe.

The output from DAOPHOT was processed to minimize contamination from bad pixels, cosmic rays, and field edges.  Stars were removed from the list based on proximity to a bad column, a high $\chi^2$ value, or a sharpness greater than 2$\sigma$ from the mean at its magnitude.  Finally, stars were matched between the many exposures for each filter, and a $UBV$ photometry list and a $BV$ photometry list created.

Photometry was compared with the Balona \& Jerzykiewicz (1993) $UBV$ CCD photometry near the NGC 2004 field and the Lucke (1972) $BV$ photoelectric photometry of LH 77 (two of the Constellation III fields).  Our values were consistent with the Balona \& Jerzykiewicz (1993) results, but calibration problems on the first night (in which the SE field was observed) forced us to use the Lucke (1972) photoelectric photometry as calibration, with offsets of $\Delta V=0.010\pm0.009$ and $\Delta B=0.031\pm0.011$.

$V, B-V$ CMDs for both fields are shown in Figure \ref{fig-4-cmds}.  Unfortunately no data was available for foreground star subtraction (the fields studied here were taken as the background fields of Dolphin \& Hunter 1997), meaning that foreground contamination cannot be corrected for.  A calculation using field star densities from Ratnatunga \& Bahcall (1985) indicates that the combined CMD would have approximately 600 Galactic foreground stars with $V$ less than 21, much less than the 21000 stars we observe.  Thus the overall CMD is very clean, with only an estimated 3\% of stars being Galactic.  However, the red giant branch, defined by $V$ brighter than 19 and $B-V$ redder than 0.8, contains 1710 observed stars and an estimated 125 foreground stars for a contamination fraction of over 7\%.  Contamination at this level will certainly affect the quality of the fits, but because the Galctic stars will not fall along LMC isochrones, the primary effect should be to increase the $\chi^2$ of the fits rather than altering the solved star formation history.

\begin{figure}
%\plotone{fig4.cmds.eps}
%\vspace{9.2cm}
\caption{Field $B-V, V$ Colour-Magnitude Diagrams for NW and SE fields}
\label{fig-4-cmds}
\end{figure}

\subsection{Artificial Stars}

Artificial star tests were made on all fields, and are described in detail by Dolphin \& Hunter (1997).  The DAOPHOT reduction of the fields with artificial stars added was identical to the original photometry, so that the artificial star photometry would be of comparable quality to the stars.  The primary difference between the artificial and observed stars is that the artificial stars, naturally, had no PSF-fitting errors.

Figures \ref{fig-4-secomp} and \ref{fig-4-nwcomp} show the completeness levels in the three filters, while Figures \ref{fig-4-sesig} and \ref{fig-4-nwsig} show the photometric error as determined by the artificial star tests.  The measured errors from the artificial star tests were significantly higher than the DAOPHOT uncertainties, which is expected for crowded field work.
The levels at which completeness in the SE field drops below half are $B$=22.3 and $V$=21.8; the levels at which photometric uncertainty (including crowding errors) increases above 0.2 magnitudes are $B$=19.5 and $V$=19.3.  Values for the NW field are approximately 0.5 magnitudes less.

\begin{figure}
%\plotone{fig4.secomp.eps}
%\vspace{9.2cm}
\caption{SE field completeness, as determined by the fraction of recovered artificial stars}
\label{fig-4-secomp}
\end{figure}

\begin{figure}
%\plotone{fig4.nwcomp.eps}
%\vspace{9.2cm}
\caption{NW field completeness, as determined by the fraction of recovered artificial stars}
\label{fig-4-nwcomp}
\end{figure}

\begin{figure}
%\plotone{fig4.sesig.eps}
%\vspace{9.2cm}
\caption{SE field photometry uncertainties, as determined by the differences in recovered and input magnitudes of artificial stars}
\label{fig-4-sesig}
\end{figure}

\begin{figure}
%\plotone{fig4.nwsig.eps}
%\vspace{9.2cm}
\caption{NW field photometry uncertainties, as determined by the differences in recovered and input magnitudes of artificial stars}
\label{fig-4-nwsig}
\end{figure}

\section{ANALYSIS}

\subsection{Star Formation History Solutions}

For all CMD analyses in this paper, the Padova isochrones (Girardi et al. 1996; Fagotto et al. 1994) were used.  These models include later (post-helium flash) phases of stellar evolution, despite the large uncertainties in those phases, providing completeness at the cost of accuracy.  For the purpose of matching the entire observed CMD with synthetic data, however, it is far better to use models with slightly incorrect AGB tracks than ones with no AGB tracks.

A standard least-$\chi^2$ fit can be easily thrown off when dealing with common observational errors, most notably the presence of observed ``stars'' (cosmic rays, noise, binary stars, Galactic foreground contamination, etc) in regions of the CMD where the models do not predict any stars.  Such a case will give a number of expected stars of $0 \pm 0$ in that region of the CMD, which will clearly make it the most important part of the fit.  To compensate, a small constant value was added to the $\sigma^2$ denominators of the $\chi^2$ calculations.  In addition, the fit parameter was modified so that any points with worse than a 3 $sigma$ error would contribute $3 \times \chi$ to the fit, rather than $\chi^2$.  Both of these modifications were made so that unfittable regions of the CMD (where no stars were predicted but stars were observed, or where models poorly reproduce the data) would not dominate the fit.

\subsection{Era of Initial Star Formation of the LMC}

The time of initial star formation in this field is an essential parameter to calculate before determining the galaxy's star formation history.  This is because the evolutionary models reproduce neither the horizontal branch nor the red clump structures very well, and thus a blind solution using the models can easily be thrown off in dealing with these old populations.  In the case of LMC, doing so could cause the models to attempt to recreate the tight red clump as a red horizontal branch.  But clearly the CMD has no pronounced horizontal feature, which sets the upper limit on large amounts of star formation to $\sim$12 Gyr ago.  This limit has been adopted into the models below, which have a 12 Gyr maxmimum age.

To determine the age of initial star formation, a set of history calculations was made using the method described in Dolphin (1997) that resolved the age of initial star formation, with the distance, extinction, and enrichment history allowed to vary.  The Padova isochrones (Girardi et al. 1996; Fagotto et al. 1994) were used for this and all other solutions in this paper.  Of the possible starting points, the oldest (12 Gyr) gave the best fits and is used below.  The question of whether the galaxy actually had \textit{no} star formation until 12 Gyr ago, or whether the galaxy had a non-zero but very small star formation rate cannot be solved from this data, as there are stars where the horizontal branch would be expected, but no horizontal feature is seen in the CMD.  Regardless of whether or not the galaxy waited until 12 Gyr ago to form any stars, it waited until that point to form a measurable number of stars.

\subsection{Global Star Formation History and Enrichment}

In order to allow for the fullest use of the CMD-matching algorithm used to determine the star formation histories, a solution for the distance and extinction to the fields was made in addition to the solution for the star formation history.  To do this, many star formation history solutions were made, each with a different combination of those parameters.  The best fits were then combined, with standard deviations of the parameters of the best fits taken as the uncertainties in the solutions.  Monte Carlo simulations were also run to attempt to determine the uncertainties inherent in the fitting procedure itself, and the uncertainties from that were added in quadrature to determine the uncertainties.  Thus the quoted standard errors are the total uncertainty in the measurements, excluding effects of uncertainties in the models.  The solution was made using 22576 stars with $B$ and $V$ between 15.5 and 21.5.  The determined extinction to the two fields was A$_V$ = 0.30 $\pm$ 0.05 and the distance modulus was $(m-M)_0 = 18.41 \pm 0.04$), which are discussed in the summary.  There is no evidence for differential reddening in the fields studied.  (Note that the broad swath of stars running diagonally from the top of the CMD to the red edge is due to Galactic foreground stars, which for reasons given above could not be subtracted.)

The star formation history with formal errors is shown in Table \ref{tab-4-oldsfh} and Figure \ref{fig-4-oldsfh}.  Average star formation rates for each era are given assuming a Salpeter IMF with cutoffs at 120 and 0.1 $M_\odot$, which is consistent with the Holtzman et al. (1997) results.  An IMF error would simply have the result of changing the ratio of old to young stars, but would not qualitatively alter the two-burst structure observed.

\begin{table*}
\caption{Star Formation History of Two LMC Fields}
\begin{tabular}{crr}
Age (Gyr) & SFR ($10^{-5} M_\odot$/yr) & [Fe/H] \\
\hline
 0$-$1   & 35 $\pm$  2 & -0.38 $\pm$ 0.10 \\
 1$-$2.5 & 26 $\pm$  4 & -0.54 $\pm$ 0.10 \\
 2.5$-$5 &  2 $\pm$  3 & -0.83 $\pm$ 0.10 \\
 5$-$7   &  3 $\pm$  4 & -1.06 $\pm$ 0.09 \\
 7$-$9   & 24 $\pm$  7 & -1.32 $\pm$ 0.09 \\
 9$-$12  & 44 $\pm$  6 & -1.63 $\pm$ 0.10 \\
\end{tabular}
\label{tab-4-oldsfh}
\end{table*}

\begin{figure}
%\plotone{fig4.oldsfh.eps}
%\vspace{9.2cm}
\caption{Star Formation History of Two LMC Fields}
\label{fig-4-oldsfh}
\end{figure}

\begin{figure}
%\plotone{fig4.recon1.eps}
%\vspace{9.2cm}
\caption{Reconstructed Colour-Magnitude Diagram of Two LMC Fields}
\label{fig-4-recon1}
\end{figure}

A synthetic CMD for the region reconstructed from the models, artificial star results, and determined star formation history is shown in Figure \ref{fig-4-recon1}b, with the observed data shown in Figure \ref{fig-4-recon1}a.  The artificial star tests for this data set clearly do not account for all observational effects, as the main sequence sharpness, empty region between the main sequence and blue loops, and other effects occur in the reconstructed CMD.  This could be a result of any combination of cosmic rays, PSF fitting errors (see above), bad pixels, etc.  Additionally, the reconstructed CMD assumes a single metallicity for each age of stars, so a broader main sequence could be constructed with a spread in the age-metallicity relation.

A small red horizontal branch appears in the reconstructed data, indicating that either the time of initial star formation or the initial metallicity could be incorrect, or that the isochrones are incorrect.  This is of less concern, as the theoretical models do not reproduce the red clump or the horizontal branch terribly well in any case.  For the most part, the RGB is well-reproduced.  The split RGB seen in the reconstructed CMD is an artifact of assuming that all stars in an age bin (ie, 9 to 12 Gyr old) have the same metallicity.  A continuous age-metallicity relation would eliminate this split.

On the main sequence, the observed data appears to cut off near $V$ of 16, while the reconstructed main sequence extends well above that.  This is a result of the chosen binning and assumption of constant star formation rate within each bin.  In this case, the youngest bin was 0 to 200 Myr old, meaning that a main sequence turnoff between $M_V$ of -2 and -3 is irreproducible with the chosen binning.  The young stellar population will be dealt with more thoroughly (with higher resolution solutions) in the following section.

Because of the large amount of foreground contamination (for example the Sun, at 1 kpc, would fall squarely in the blue loop region), comparison between observed and reconstructed CMDs in the blue loop region will not be fruitful.  Ideally, data would have been taken outside the LMC to sample foreground stars, but the data used for this paper was taken as a background field for the Constellation III study (Dolphin \& Hunter 1998).

\begin{figure}
%\plotone{fig4.compare.ps}
%\vspace{9.2cm}
\caption{Comparison of Observed and Reconstructed CMDs.  Upper left is the binned observed CMD; upper right is the binned reconstructed CMD.  The lower left shows the subtracted diagram, while the lower right shows the residual divided by the uncertainty at each location.  See text for explanation.}
\label{fig-4-comp}
\end{figure}

\begin{figure}
%\plotone{fig4.compare2.ps}
%\vspace{9.2cm}
\caption{Comparison of Observed and Reconstructed CMDs.  Same as Figure \ref{fig-4-comp}, but at twice the resolution.}
\label{fig-4-comp2}
\end{figure}

A more detailed comparison is shown in Figures \ref{fig-4-comp} and  \ref{fig-4-comp2}, which show the binned CMDs that were used for the solution (0.1 magnitude resolution in $B-V$ and 0.3 magnitude resolution in $B$ for Figure \ref{fig-4-comp}; 0.05 $B-V$ and 0.15 $B$ for Figure \ref{fig-4-comp2}), as well as the residuals and the fit parameter values.  Magnitude limits in the plots are $15.5 < B < 20.9$ and $-0.4 < V-I < 2.4$.
The subtracted CMD is shown at twice the contrast level of the two CMDs (darker being undersubtracted and lighter oversubtracted), while the Hess diagrams are shown on a scale from -9 $\sigma$ to +9 $\sigma$.

The fitting procedure was able to fit the main sequence reasonably well, with all points in the main sequence area being fit within 3 $\sigma$.  A broader theoretical main sequence (either from larger photometric errors or an age-metallicity spread) would have eliminated most of the larger errors in the main sequence fit.  The upper red giant branch was also well-fit, with nearly all points being fit within 2 $\sigma$.

By far the worst-fit part of the CMD was the region with the red clump, and horizontal branch, which contained the 9 $\sigma$ errors in the fitting.  Given the limitations of the isochrones used, these errors are not surprising.  Since the robust fit parameter used here was less sensitive to this than a traditional $\chi^2$ fit would have been, this had a minimal effect on the overall star formation history solutions.

The star formation in the LMC apparently began with a significant episode, which accounted for about half of the LMC's total star formation (by mass of stars formed) in the first 3 Gyr.  After that, the star formation rate appears to have slowly declined from 11 Gyr ago until about 2.5 Gyr ago.  During more recent times, beginning between 1.5 and 2.5 Gyr ago, the LMC has undergone a second large episode of star formation, which is studied in greater detail in the following section.  This history appears to support the recent HST-based results (Holtzman et al. 1997; Stappers et al. 1997; Geha et al. 1998; and Olsen 1999), with the recent burst beginning about 2.5 Gyr ago.

The metallicity of the LMC's oldest population of stars is [Fe/H] = -1.63 $\pm$ 0.10.  This is consistent with Holtzman et al. (1997)'s estimate of [Fe/H] = -1.7 for an old RGB.  It has climbed very steadily since then, to its present value of -0.38 $\pm$ 0.10, consistent with the findings of Holtzman et al. (1997) and Geha et al. (1998).

\subsection{Recent Star Formation History}

With more recent star formation, it is beneficial to study the two fields separately, as recent star formation in the two fields is likely to be quite different.  Specifically, the fields are on either side of LMC Constellation III, a $\sim$1 kpc diameter superbubble where significant amounts of recent star formation have taken place (Dolphin \& Hunter 1998).  Although the Constellation III event happened very recently (in the past $\sim$20 Myr), it is possible that the recent star formation histories in these two fields have been quite different over a longer period.

Using the distance and extinction determined in the previous section, solutions for the two fields were run that had higher resolution for recent star formation rates and events.  The number of stars in each field was 10354 in NW and 12222 in SE.  Resulting star formation rates are given in Table \ref{tab-4-newsfh} and Figure \ref{fig-4-newsfh}.

\begin{table*}
\caption{Recent Star Formation History of LMC Fields}
\begin{tabular}{crr}
Age (Gyr) & NW SFR ($10^{-5} M_\odot$/yr) & SE SFR ($10^{-5} M_\odot$/yr) \\
\hline
 0.0$-$0.1   & 14 $\pm$  5 & 26 $\pm$ 6 \\
 0.1$-$0.2   & 22 $\pm$  3 & 38 $\pm$ 4 \\
 0.2$-$0.4   & 11 $\pm$  2 & 17 $\pm$ 2 \\
 0.4$-$0.6   & 13 $\pm$  3 & 14 $\pm$ 3 \\
 0.6$-$1.0   & 19 $\pm$  2 & 17 $\pm$ 1 \\
 1.0$-$1.5   & 15 $\pm$  2 & 14 $\pm$ 2 \\
 1.5$-$2.5   & 10 $\pm$  2 & 15 $\pm$ 2 \\
 older ages  & 12 $\pm$  1 &  9 $\pm$ 1 \\
\end{tabular}
\label{tab-4-newsfh}
\end{table*}

\begin{figure}
%\plotone{fig4.newsfh.eps}
%\vspace{9.2cm}
\caption{Recent Star Formation History of Two LMC Fields}
\label{fig-4-newsfh}
\end{figure}

\begin{figure}
%\plotone{fig4.recon2.eps}
%\vspace{9.2cm}
\caption{Separately Reconstructed Colour-Magnitude Diagrams of Two LMC Fields}
\label{fig-4-recon2}
\end{figure}

Reconstructed CMDs of the two fields are shown in Figure \ref{fig-4-recon2}, which can be compared with the observed data in Figure \ref{fig-4-cmds}.  Aside from the fitting problems mentioned above in relation to the combined CMD, the agreement is excellent.  Again, the split red giant branch is a result of the assumption that the metallicity is constant over each age bin.  Thus there is a jump between [Fe/H] of -1.63 $\pm$0.10 and -1.32 $\pm$0.09 for the 9-12 and 7-8 Gyr old bins respectively, when in reality the metallicity would have gradually increased.  However, the total width of the red giant branch is well-reconstructed.

The star formation histories of the two fields are remarkably similar, and are consistent with each other until about 200 Myr ago.  From 200 to 2500 Myr ago, the average star formation rate in the two fields is $1.4 \times 10^{-4} M_\odot$/yr, $30\%$ higher than the average star formation rate between 2.5 and 12 Gyr ago.  The NW field, overall, is consistent within 2.5 $\sigma$ of a constant star formation rate of $1.4 \times 10^{-4} M_\odot$/yr over the recent 2500 Myr episode.

The SE field, however, has shown a very strong increase in star formation activity over the past 200 Myr, as shown in Table \ref{tab-4-newsfh} (as well as with the casual observation that the SE field's CMD shows many more upper main sequence stars than does the NW field's CMD).  The star formation rate during this period is consistent with a constant value of $3.4 \times 10^{-4} M_\odot$/yr, a 140\% increase over the 2500 Myr average rate.

With both fields having consistent star formation rates for each bin until very recently, it follows that whatever conditions caused this recent (2500 Myr) outburst of star formation happened over a large scale.  Given their physical separation of $\sim$2 kpc, it seems unlikely that a single localized event could have triggered the large amount of star formation seen here.  Also, with other studies of the LMC also finding a recent burst, there must have been some large change of environment that triggered an era of star formation throughout the galaxy.  The more recent (200 Myr) event was localized, only affecting the SE field.

\section{SUMMARY}

Ground-based UBV photometry of two fields in the northern disk of the LMC has been presented.  A distance modulus of 18.41 $\pm$ 0.04 and an extinction of 0.30 $\pm$ 0.05 have been calculated.  The distance is smaller than the SN 1987A distance of 18.55 $\pm$ 0.05 (Panagia 1998), but consistent with the red clump distance of 18.36 $\pm$ 0.17 (Cole 1998).  The extinction agrees with the average extinction of $A_V = 0.40 \pm 0.11$ found in the two fields by Dolphin \& Hunter (1998).  The LMC's oldest significant population appears to have a metallicity [Fe/H] of -1.63 $\pm$ 0.10, with an age of initial star formation of roughly 12 Gyr to account for the lack of a horizontal branch in the observed data.  About half of the LMC's stars (by mass) formed by 9 Gyr ago.

After this initial event, the star formation tapered off slowly, until a recent burst beginning about 2.5 Gyr ago.  This result is consistent with most other LMC star formation history works, as it follows the on-off-on star formation history.  The metallicity has been constantly rising, and is currently at [Fe/H] = -0.38 $\pm$ 0.10, consistent with the findings of Geha et al. (1998) and Holtzman et al. (1997).  This recent episode dramatically strengthened in the SE field about 200 Myr ago, but the NW field is consistent with a constant star formation rate for the past 2.5 Gyr.

In summary:
\begin{itemize}
\item The distance modulus for this portion of the LMC was determined photometrically to be $(m-M)_0 = 18.41 \pm 0.04$.  This is shorter than the SN 1987A distance of 18.55 $\pm$ 0.05 (Panagia 1998), but consistent with a red clump distance of 18.36 $\pm$ 0.17 (Cole 1998).
\item The oldest significant star formation in the LMC occurred about 12 Gyr ago, with about half of the LMC's star formation occuring in the first 3 Gyr.  After decreasing star formation until about 2.5 Gyr ago, a recent burst has rekindled the LMC.  In agreement with the recent HST-based results, the burst appears to have begun in the 2-3 Gyr range (rather than the 3-5 Gyr range favored by previous ground-based studies).
\item The metallicity of this part of the LMC has increased from an initial value of -1.63 $\pm$ 0.10 to a present value of -0.38 $\pm$ 0.10.
\item The two fields studied show very similar star formation histories over the past 2.5 Gyr, but have deviated significantly from one another during the past 200 Myr.  Given the separation of $\sim$2 kpc, the environmental change that caused this burst must have occurred over a very large, perhaps global scale in the LMC, with the recent ($<$200 Myr) star formation beginning to show regional differences.
\end{itemize}

\section*{Acknowledgments}

I am indebted to Deidre Hunter for obtaining the data used in this paper.

\bsp

\label{lastpage}

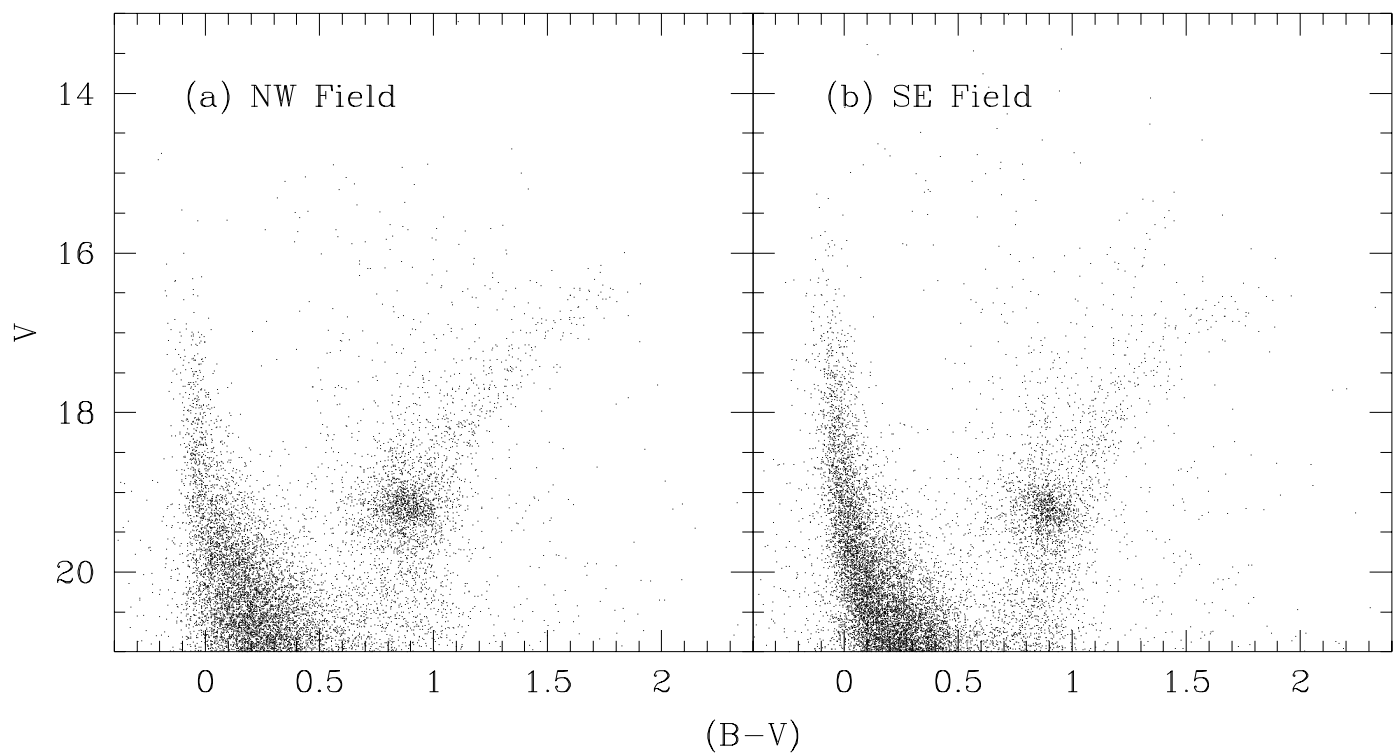
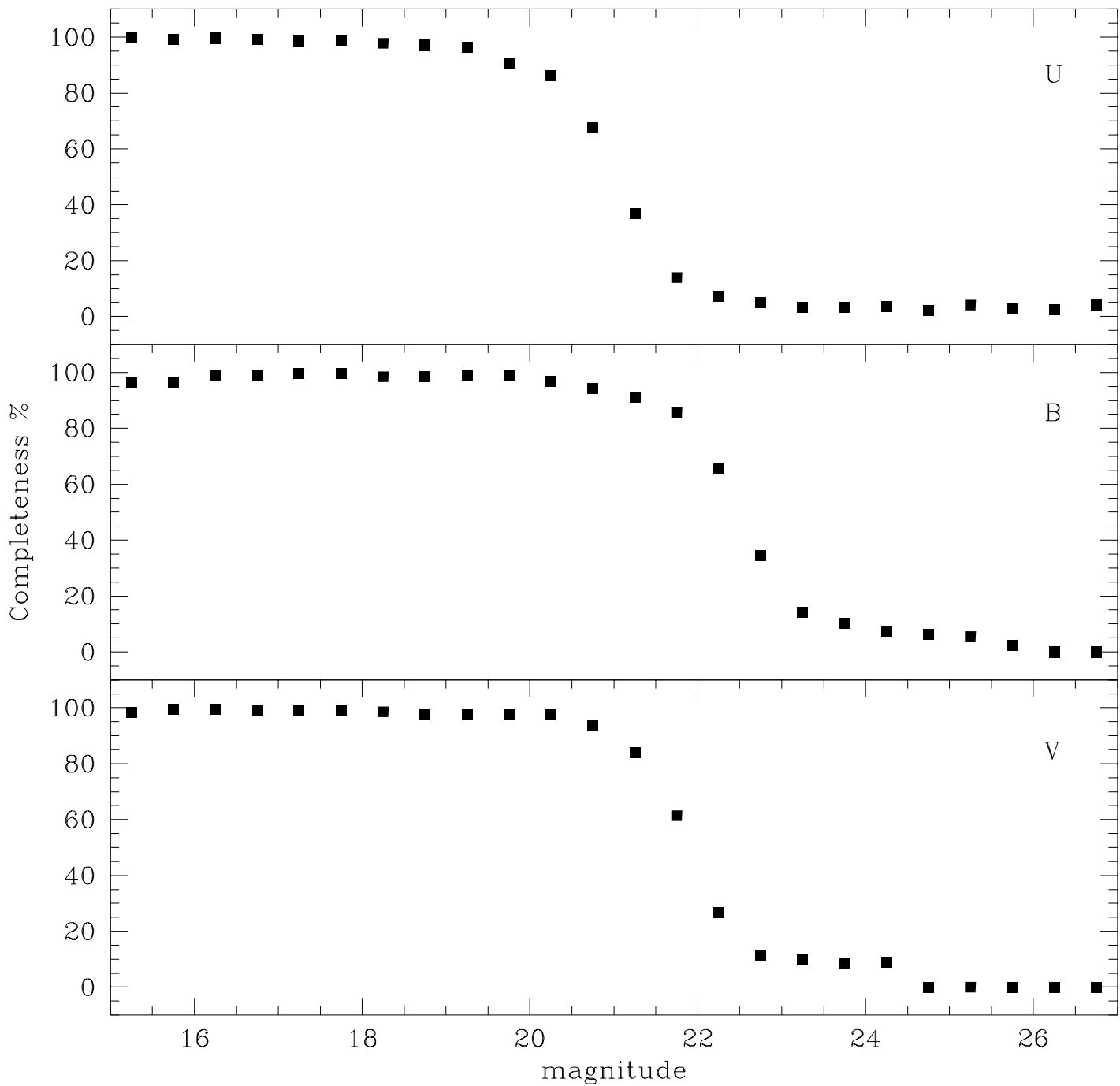
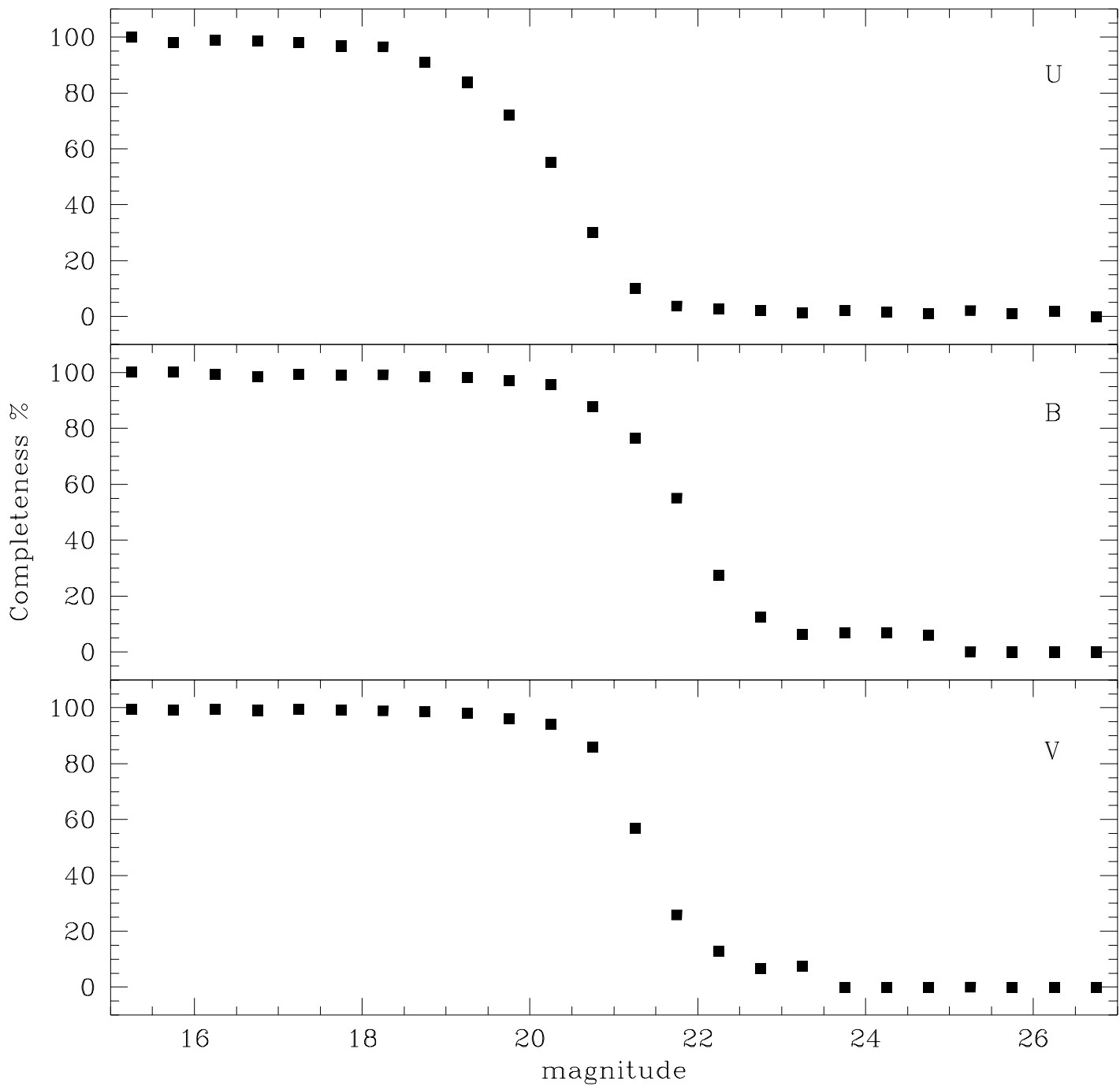
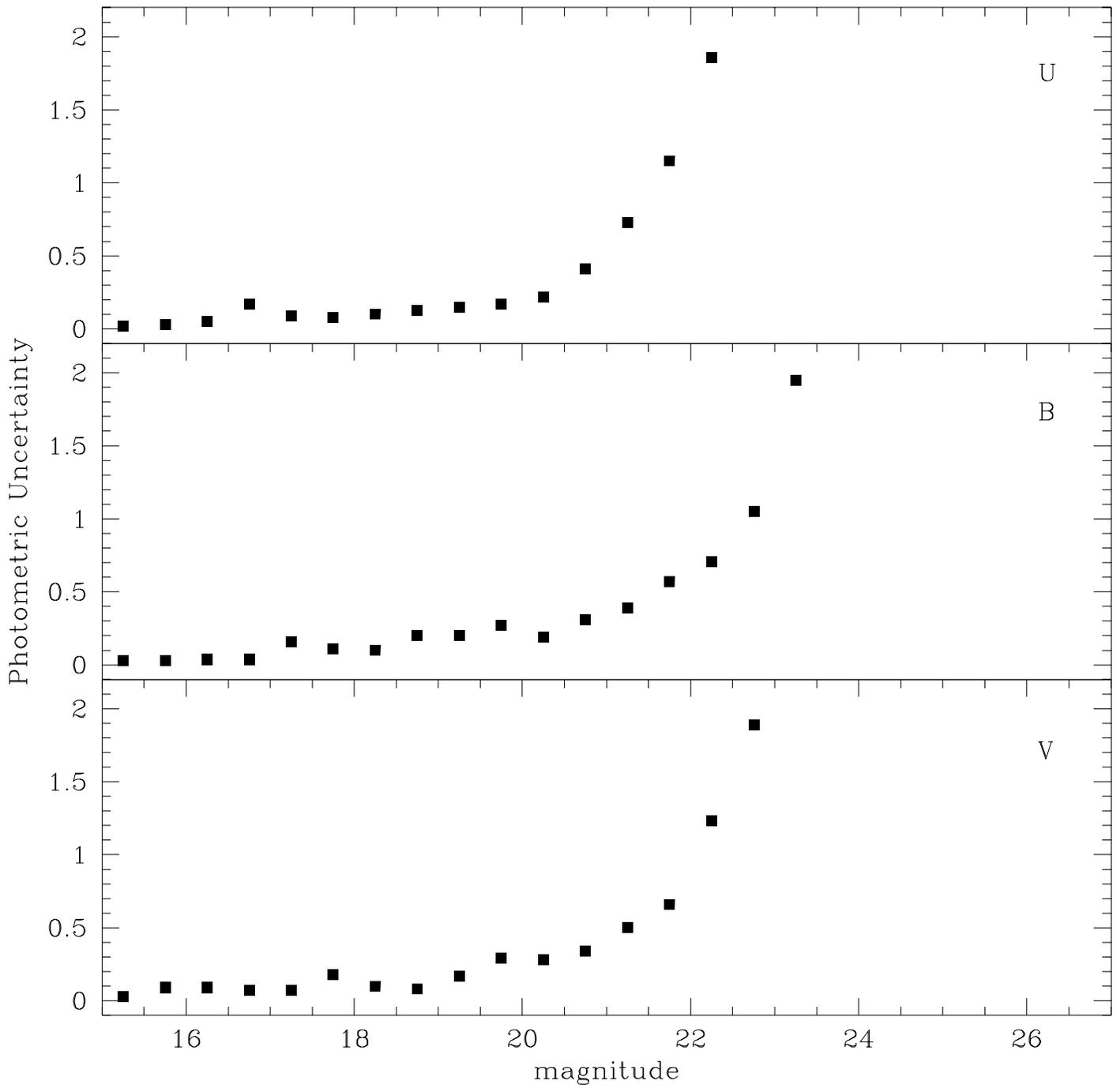
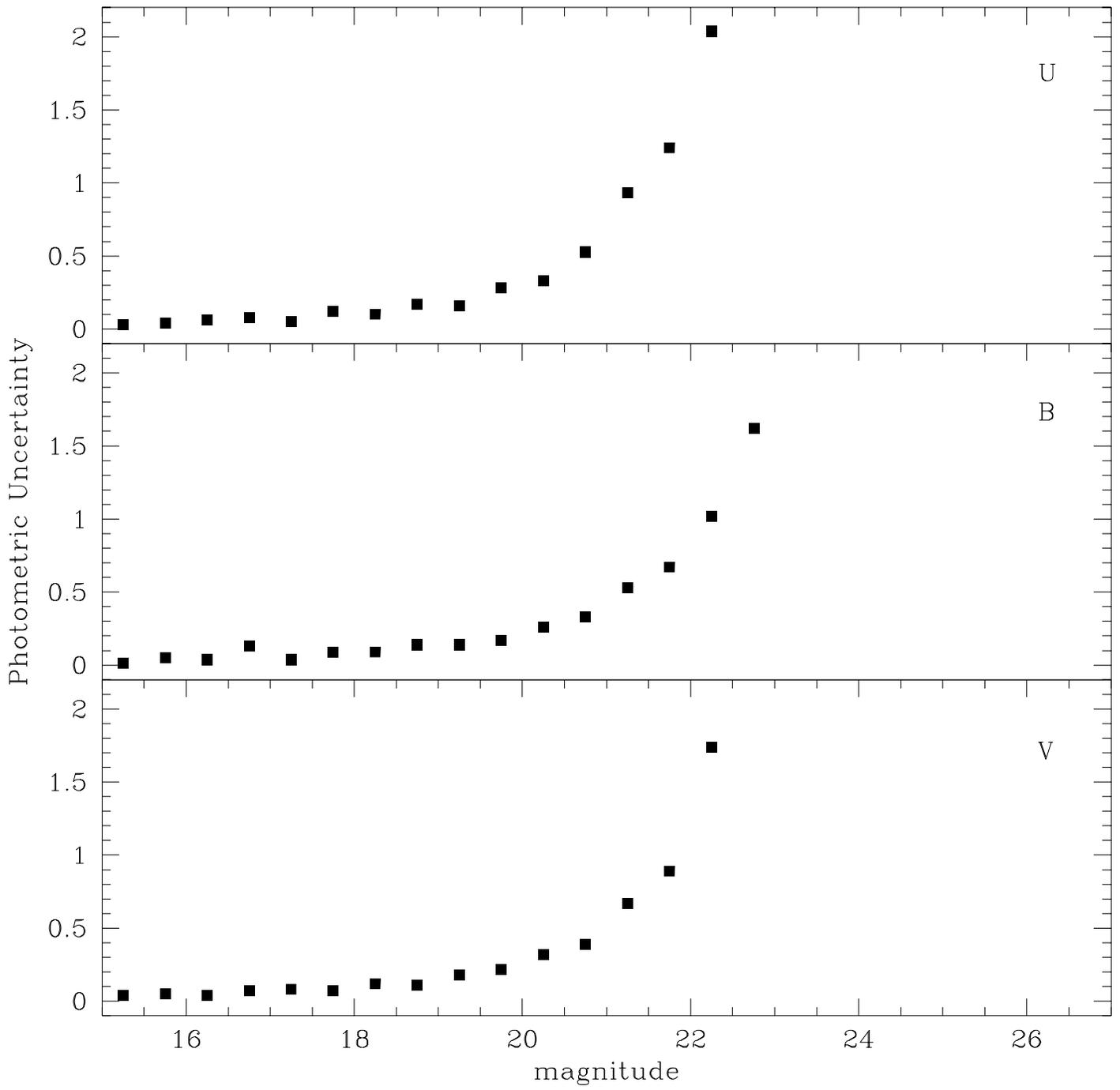
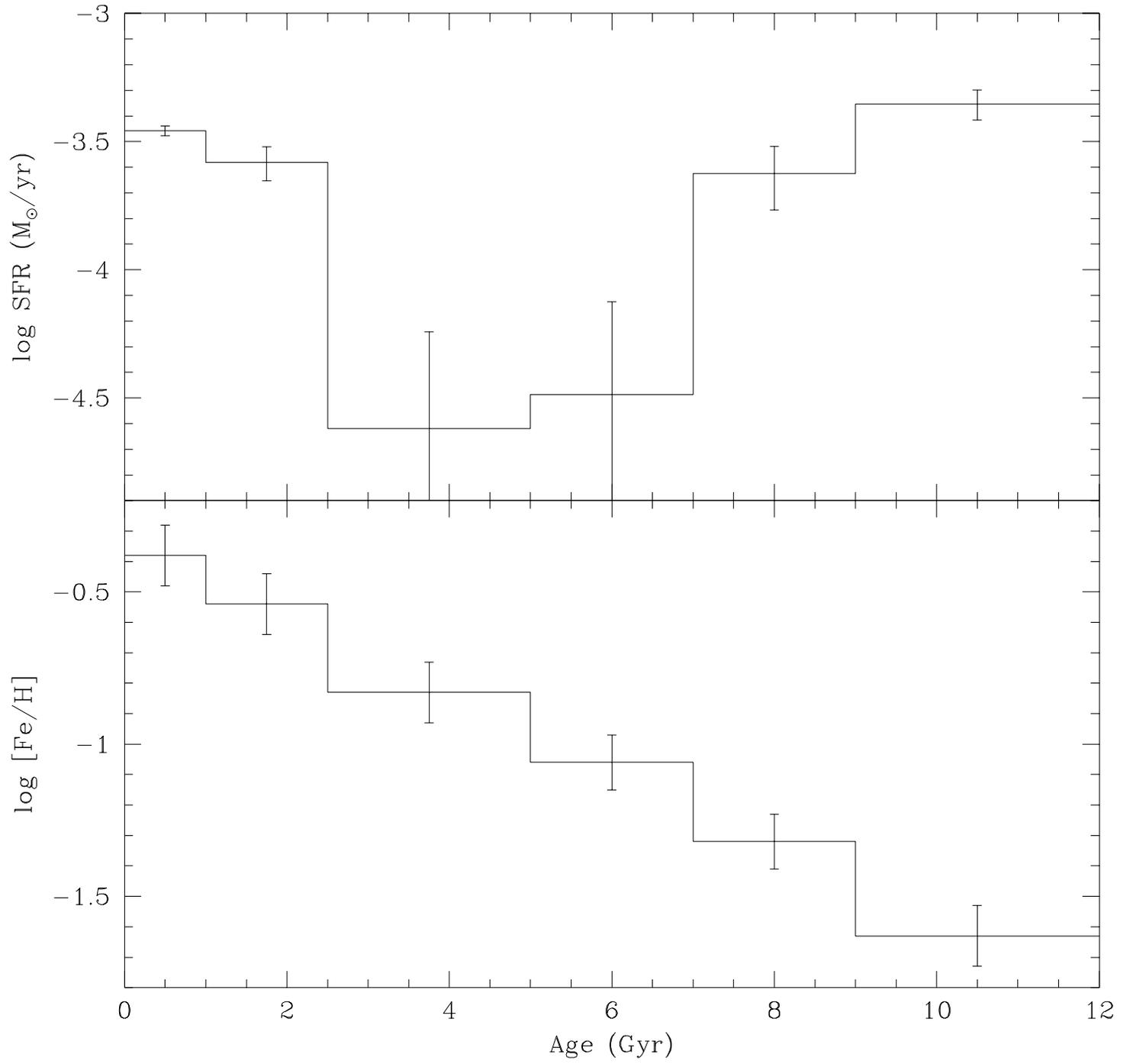
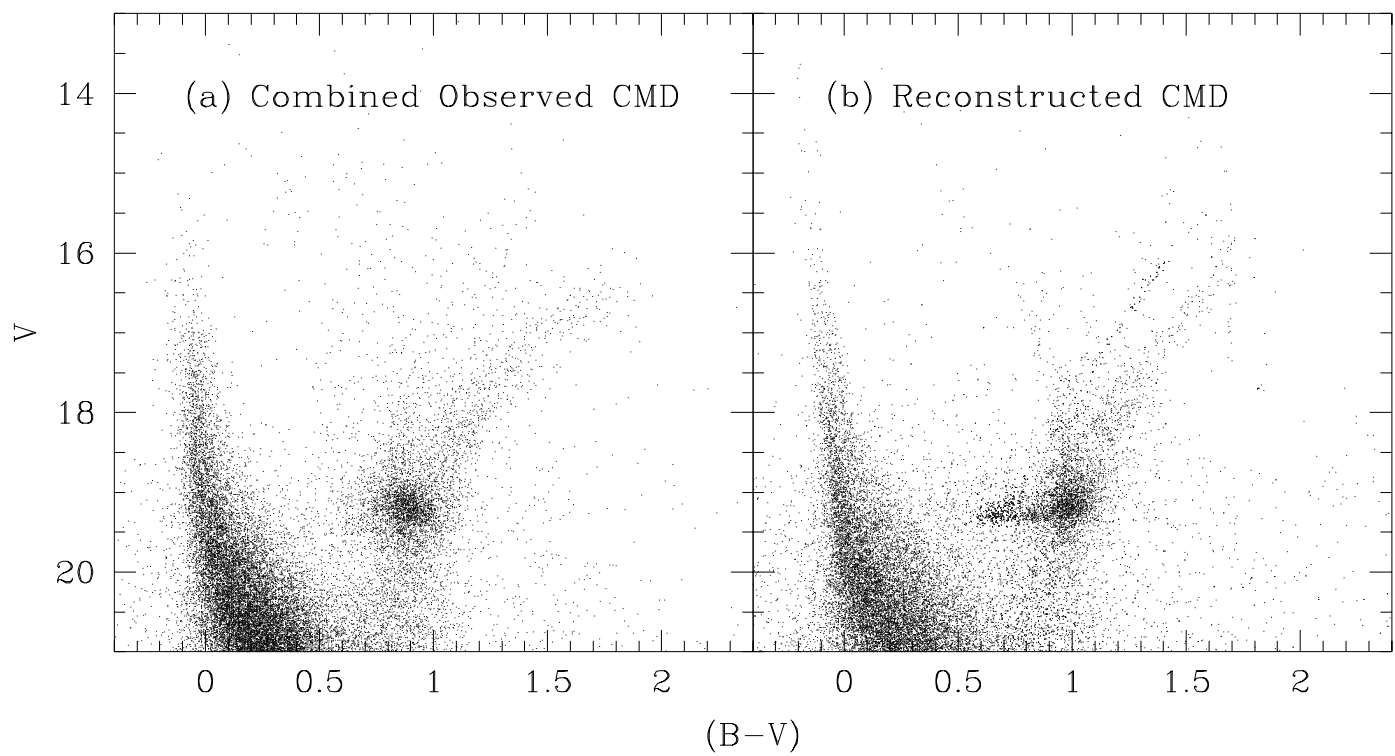
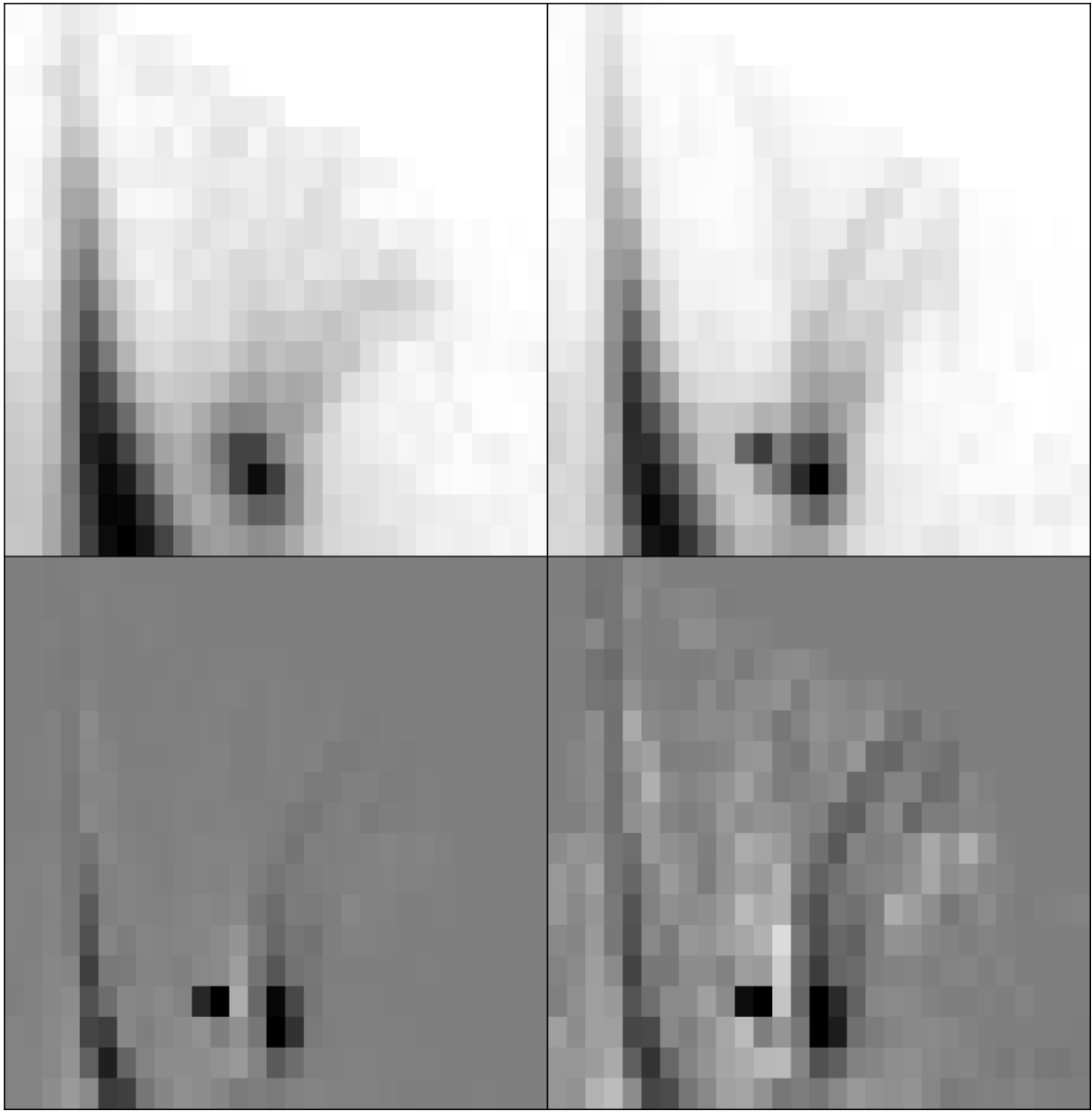
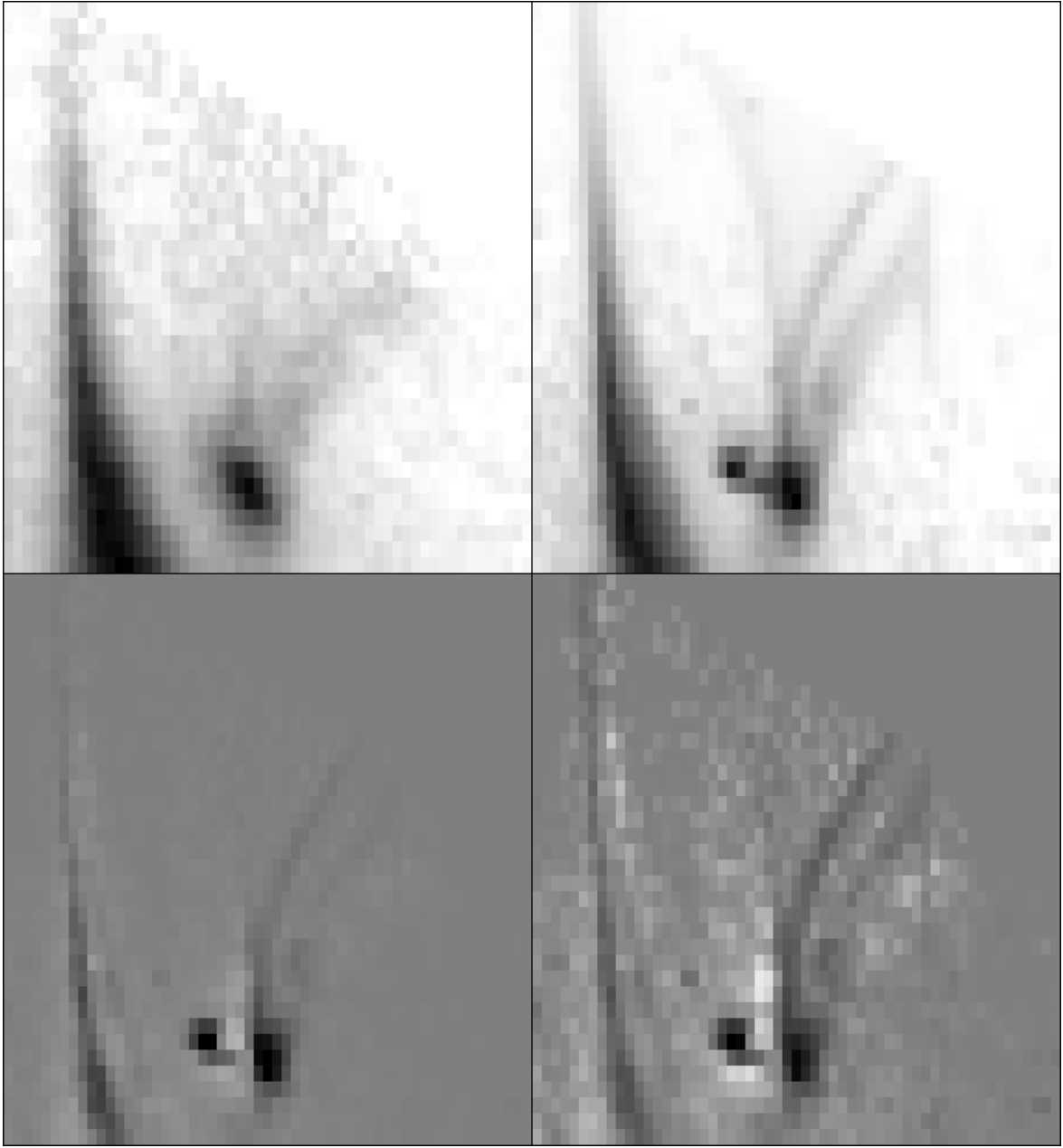
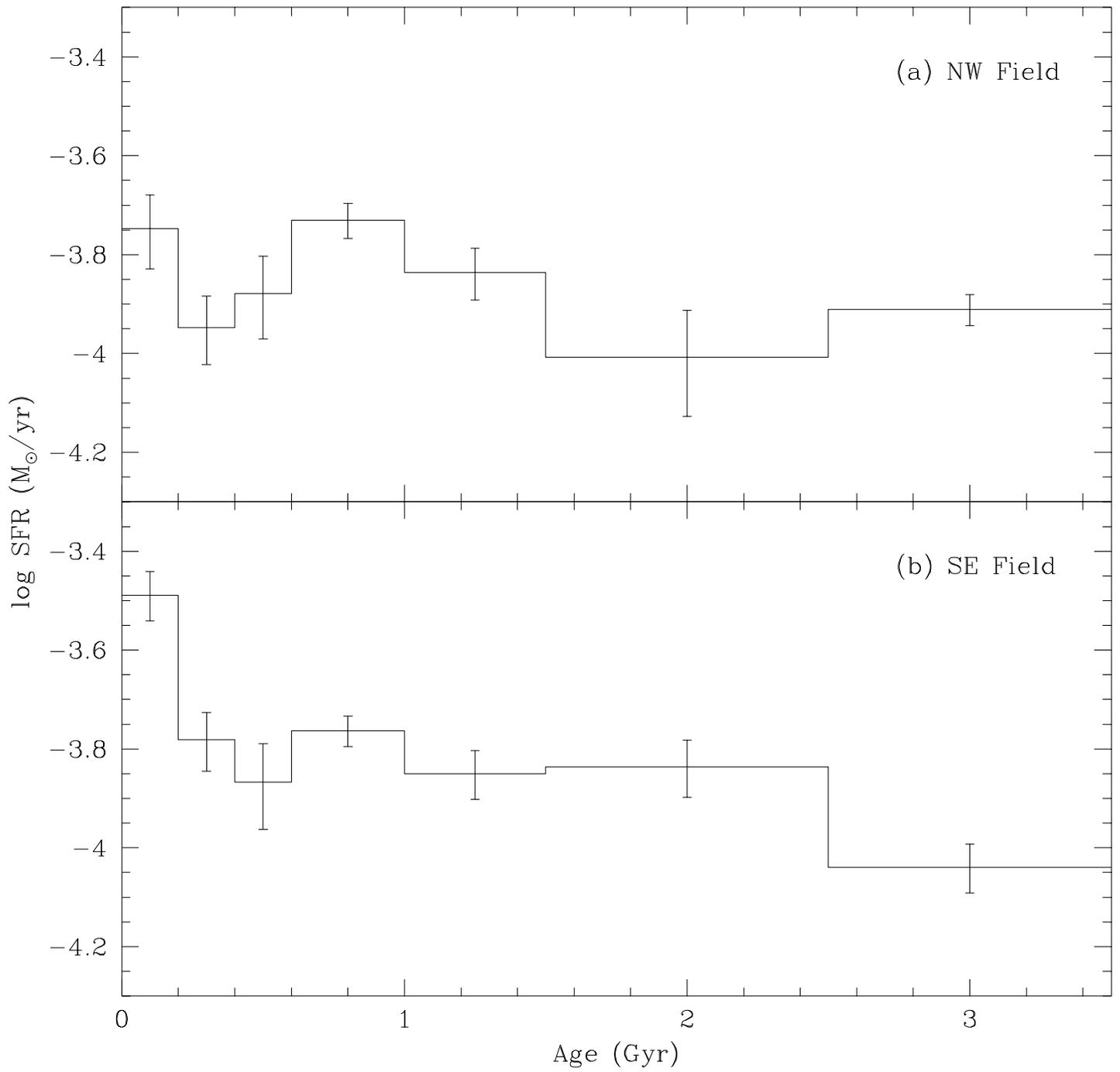
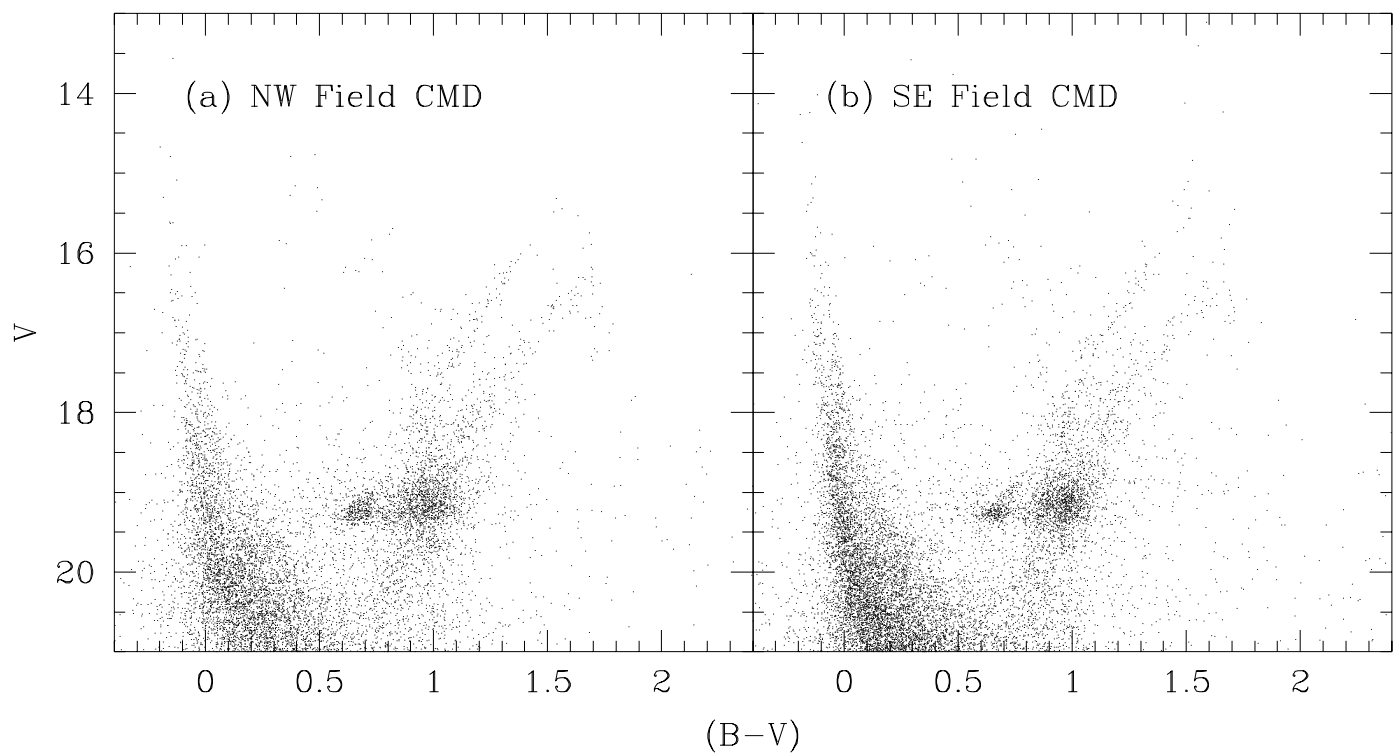
\end{document}